%% file: main.tex
\documentclass[sigconf,nonacm]{acmart}

\usepackage{booktabs}
\usepackage{array}
\usepackage{graphicx}
\usepackage{microtype}
\usepackage{xcolor}

\renewcommand\footnotetextcopyrightpermission[1]{}
\newcommand{\loss}{L_F}
\newcommand{\benefit}{B_F}

\title{When Does Permutation Instability Generalize? Independent-View Validation for Listwise LLM Reranking}

\author{Wenzhang Du}
\affiliation{%
  \institution{Independent Researcher}
  \country{China}
}

\begin{abstract}
Listwise language-model rerankers often disagree across equivalent candidate
permutations. Finite instability diagnostics are therefore used to motivate
additional sampling, aggregation, or selective computation. But an association
with a validation statistic that reuses the probe views need not isolate
predictive information about \emph{unseen} permutations. Shared measurements
can induce classical part--whole association. We study how this affects claims
that a finite-view instability score predicts unseen permutations. We derive
the exact finite-view decomposition and prospectively compare zero, one, and
two reused views, including a fully disjoint four-view target.

The study covers two pinned 7B model families and two recommendation datasets,
with controlled lists for signed offline analysis and untouched retriever lists
for target-free replication. On the four controlled blocks, fully disjoint
correlations are weak or heterogeneous ($-0.061$--$0.281$), whereas reusing both
probe views yields $0.600$--$0.718$; all paired contrasts are large
($0.436$--$0.661$) and Holm-significant. The overlap effect is positive in all
four untouched-list blocks. Increasing the probe from two to four views clearly
improves disjoint reliability in only one block. Moreover, the probe predicts
aggregation-movement magnitude ($\rho=0.142$--$0.426$) but not stable signed
target benefit, and 7 of 12 fixed-fraction probe-routing points are strictly
dominated at measured cost. Thus, when the intended estimand is predictive
information about unseen perturbation behavior, validation targets must be
observation-disjoint from the probe to isolate that information; signed utility
and cost-sensitive decisions remain separate questions.
\end{abstract}

\ccsdesc[500]{Information systems~Recommender systems}
\ccsdesc[500]{Information systems~Retrieval models and ranking}
\ccsdesc[300]{Computing methodologies~Machine learning approaches}
\keywords{LLM reranking, order sensitivity, measurement validity, perturbation views, selective computation}

\begin{document}
\maketitle

\section{Introduction}

Listwise language models rerank candidates by reading a serialized prompt and
emitting an ordering \cite{hou2024llmrankers,sun2023rankgpt}. The candidate set
is unordered, yet decoder-only rerankers can change scores, preferences, and
rankings when that serialization changes. Order sensitivity and positional bias
are established phenomena rather than discoveries of this work
\cite{tang2024permutation,bito2025position,bito2026preference}. The practical
question is what to do with a small number of inconsistent outputs.

Finite disagreement scores can serve several operational roles. They may
motivate averaging more permutations, choosing an invariant intervention,
falling back to a different ranker, or allocating extra inference only to
apparently difficult queries. Permutation self-consistency already aggregates
repeated rankings \cite{tang2024permutation}; RoToR selectively routes listwise
inputs \cite{yoon2025rotor}; and AcuRank adapts computation using relevance
uncertainty \cite{yoon2025acurank}. These methods make a diagnostic's
interpretation consequential: a score can be reproducible on the observations
that produced it yet fail to predict the behavior needed by a later action.

We ask a measurement question that precedes the intervention: \emph{what
evidence licenses a finite-view diagnostic as a predictor of unobserved
perturbation views?} A high association with a larger finite-view statistic is
often read as support for such extrapolation. If the larger statistic contains
the same views used by the probe, however, the evaluation changes two things at
once: it measures fresh-view behavior and reuses calculated components. The
quantity no longer isolates the extrapolative relation being claimed.

The dependence induced by shared calculated components is classical.
Part--whole correlation and mathematical coupling have long been studied
\cite{henrysson1963itemtotal,archie1981coupling,
blance2005coupling}, while circular analysis provides broader data-reuse context
rather than an exact synonym \cite{kriegeskorte2009circular}. Our contribution
is specific to a finite perturbation diagnostic assigned a fresh-view
interpretation: the validation protocol must match that estimand. For this
purpose, ``independent-view'' means that target observations are disjoint from
probe observations, not that rankings for the same query are probabilistically
independent.

Before generating the new rankings, we fixed an eight-view deterministic bank,
a two-view probe, and four-view validation targets that reuse zero, one, or both
probe observations. Target combinations are averaged within query, and a fully
disjoint four-view target evaluates fresh-view behavior. We apply this design to
two model families and two datasets, then repeat the overlap comparison on
untouched retriever lists. The design separates an inclusive descriptive
association from the predictive information that remains on observations not
used to calculate the probe.

\begin{figure*}[t]
  \centering
  \includegraphics[width=\textwidth]{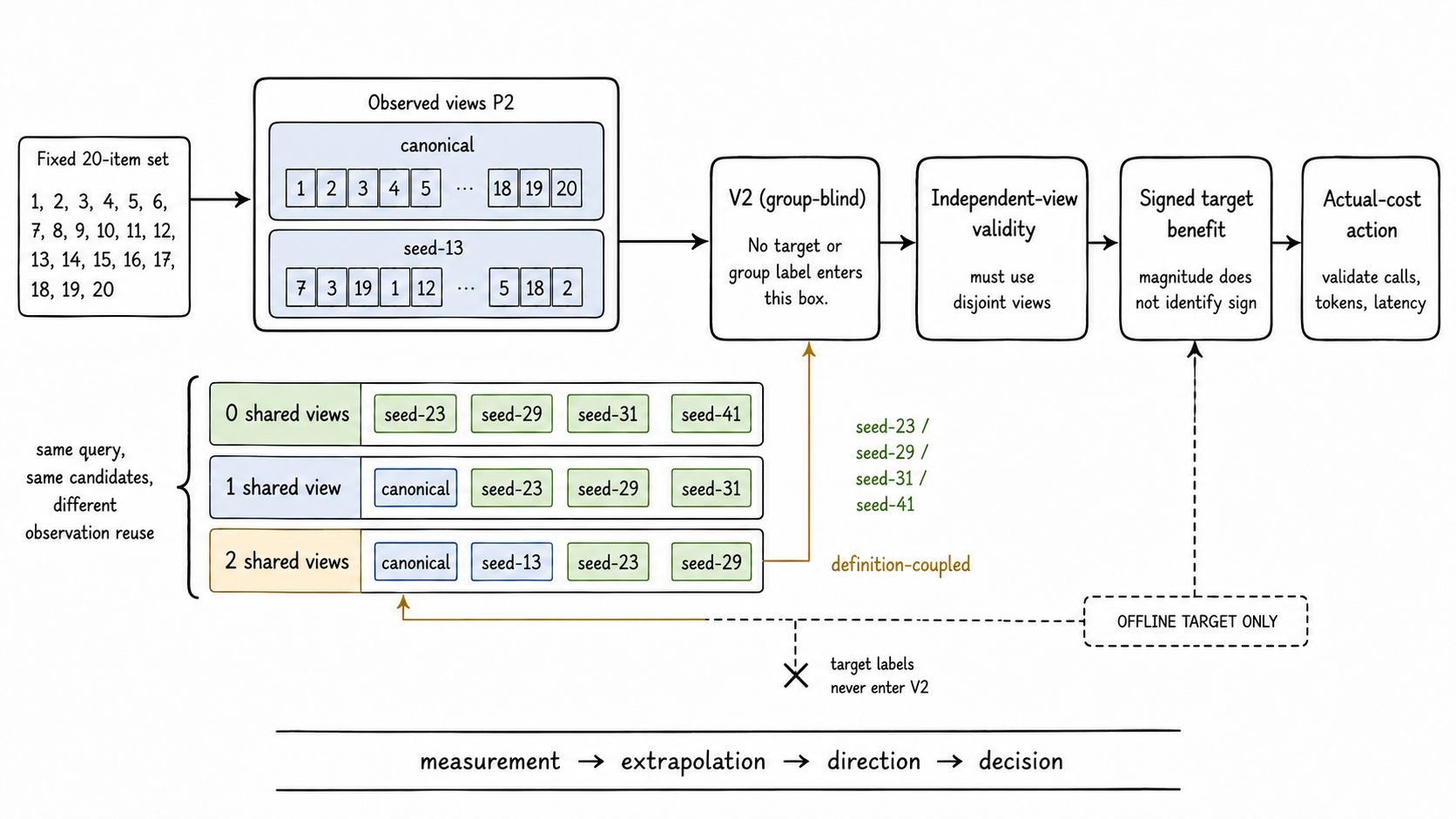}
  \Description{The validation chain begins with a two-view probe using the
  canonical and seed-13 serializations. Independent target views 23, 29, 31,
  and 41 form zero-, one-, and two-overlap targets. A diagnostic must first
  generalize to independent views, then predict signed target benefit, and
  finally improve a routing decision at measured cost. Target labels appear
  only in an offline evaluation lane.}
  \caption{Independent-view validation separates an observed two-view,
  group-blind diagnostic from three later questions. A four-view target can
  reuse zero, one, or both probe views while query and candidates stay fixed.
  Target labels enter only the offline signed outcome, never $V_2$. The diagram
  depicts the tested validity chain, not a superiority claim.}
  \label{fig:validity-chain}
\end{figure*}

Finally, extrapolation is only one rung of the decision chain. A target-blind
movement magnitude need not identify whether aggregation moves toward or away
from a signed offline target, and even a predictive score need not justify an
extra-call policy. Figure~\ref{fig:validity-chain} therefore separates
measurement, extrapolation, direction, and decision. Our contributions are:

\begin{itemize}
  \item We formulate fresh-view validity for the finite-view statistic and
  derive its exact pairwise and nested decomposition, separating algebraic
  observation reuse from empirical sign and effect size.
  \item We prospectively isolate observation reuse with a query-level
  zero/one/two-overlap experiment and a fully disjoint four-view target across
  eight fixed views, two datasets, and two model families, with untouched-list
  replication.
  \item We separate perturbation movement, signed target benefit, and
  measured-cost action, showing that movement predictiveness does not settle
  benefit or routing utility in the tested geometry.
\end{itemize}

We additionally evaluate fixed-fraction $V_2$ routing to test whether diagnostic
validity translates into a useful cost--utility policy. Together, the analyses
isolate probe--target observation reuse for a finite diagnostic with a
fresh-view interpretation and separate extrapolation, signed value, and
measured-cost action.

\section{Finite-View Validity}

\subsection{Estimand and observation sets}

For a fixed query and candidate set, a \emph{view} is one deterministic
serialization followed by one constrained model ranking. A finite probe
$V(P)$ summarizes an observed set of views $P$. The intended primary estimand
is not the inclusive association between this probe and a statistic that may
contain $P$; it is the across-query association between $V(P)$ and instability
calculated from a separate target set $T$. This estimand asks whether variation
among the observed probe views carries predictive information about unseen
views from the same pre-specified generator.

Consequently, the target observations must satisfy $P\cap T=\varnothing$ to
isolate fresh-view predictive information. This is an estimand-specific design
condition. It is not a universal rule for every prediction task: for example,
predicting externally labeled NDCG from a consistency score has a different
target and does not automatically instantiate the same part--whole structure.
Nor does observation-disjointness assert that rankings are statistically
independent after conditioning on a shared query and candidate set. It removes
the direct reuse of the measurements used to calculate the probe.

\subsection{Ranking views and exposure instability}

For one query, let $C$ be a fixed set of 20 candidates and $r_s$ the ranking
produced under deterministic serialization $s$. We map $r_s$ to a normalized
top-10 item-exposure vector $x^{(s)}$ using discount
$1/\log_2(1+\mathrm{rank})$ and zero exposure below rank 10. For $m$ views, the
implemented diagnostic is summed itemwise population variance,
\begin{equation}
 V_m=\frac{1}{m}\sum_{s=1}^{m}\|x^{(s)}-\bar{x}\|_2^2.
 \label{eq:variance}
\end{equation}
It uses no relevance, popularity, or group label.

\subsection{Exact nested-view decomposition}

The standard pairwise identity gives
\begin{equation}
 V_m=\frac{1}{m^2}\sum_{a<b}\|x^{(a)}-x^{(b)}\|_2^2,
 \qquad
 V_2=\frac14\|x^{(1)}-x^{(2)}\|_2^2.
 \label{eq:pairwise}
\end{equation}
For completeness, expand the two sides before dividing by $m^2$:
\begin{align}
 \sum_s\|x^{(s)}-\bar{x}\|^2
 &=\sum_s\|x^{(s)}\|^2-m\|\bar{x}\|^2,\\
 \sum_{a<b}\|x^{(a)}-x^{(b)}\|^2
 &=m\sum_s\|x^{(s)}\|^2-\Big\|\sum_s x^{(s)}\Big\|^2.
\end{align}
Since $\sum_sx^{(s)}=m\bar{x}$, the second expression is $m$ times the
first, proving Eq.~\ref{eq:pairwise} under the implementation's
\texttt{ddof=0}. A sample-variance normalization would instead differ by
$m/(m-1)$.

Now partition four views into two pairs $A$ and $B$, with pair means $\mu_A$
and $\mu_B$. Direct expansion yields
\begin{equation}
 V_4=\tfrac12 V_2(A)+\tfrac12 V_2(B)
      +\tfrac14\|\mu_A-\mu_B\|_2^2.
 \label{eq:nested}
\end{equation}
Equation~\ref{eq:nested} follows by the within/between decomposition. Each pair
contributes two deviations around its own mean, while each pair mean is
displaced by $(\mu_A-\mu_B)/2$ from the four-view mean. The two within sums are
$2V_2(A)$ and $2V_2(B)$; division by four gives the first two terms, and the
four between deviations give the final term.
Thus a four-view target containing the two probe views contains the probe
statistic structurally. Equivalently, the probe pair itself contributes
$\tfrac14V_2(A)$ as one explicit term in the six-pair representation. Across
queries,
\begin{align}
 \mathrm{Cov}(V_2(A),V_4)
 &=\tfrac12\mathrm{Var}(V_2(A))
 +\tfrac12\mathrm{Cov}(V_2(A),V_2(B)) \notag\\
 &\quad+\tfrac14\mathrm{Cov}(V_2(A),\|\mu_A-\mu_B\|^2).
 \label{eq:covariance}
\end{align}
The first term is mechanical, but either covariance can be negative. Therefore
the identity proves observation reuse, not a positive correlation, a universal
inflation direction, or an empirical effect size. Those require independent
evidence. As an implementation check rather than confirmation of the new
empirical object, we numerically verify Eqs.~\ref{eq:pairwise}--\ref{eq:nested}
on 1,452 legacy sealed queries to a maximum absolute error of
$2.08\times10^{-17}$.

\subsection{Unsigned motion does not identify direction}

Let $c$ and $a$ be canonical and action exposure, $t$ an external target, and
$B(c,a;t)=L(c,t)-L(a,t)$. No statistic that depends only on unsigned,
target-blind motion can generally determine the sign of $B$ without target or
alignment information. For example, choose $c=(.55,.45)$ and equally distant
actions $a_+=(.65,.35)$, $a_-=(.45,.55)$ under target $t=(.65,.35)$. Both
actions move by $\sqrt{.02}$ inside the exposure simplex, yet one has benefit
$+.10$ and the other $-.10$. The same sign reversal is realizable by swapping
head/tail items at ranks 1 and 2 under our exact discount: normalized weights
are $p_1=.2200918$ and $p_2=.1388624$, so both worlds have squared movement
$2(p_1-p_2)^2=.0131964$, while feasible target share $.20$ yields benefits
$+.0410458$ and $-.0410458$. Instability can measure displacement; direction
remains a separate empirical question.

\section{Prospective Independent-View Study}

\subsection{Study logic and freshness boundary}

Before generating any new rankings, we fixed the eight-view bank, probe and
target assignments, overlap combinations, signed outcome, bootstrap and
multiplicity procedures, routing fractions, model and dataset selection rules,
and device policy. Earlier four-view Qwen results motivated the measurement
question but were used only for discovery; they do not provide confirmatory
evidence for the independent-view conclusion.

In the controlled Qwen blocks, canonical/13/17/19 were existing measurements,
whereas 23/29/31/41 were generated only after this prospective specification.
This retains the primary probe while ensuring that its four target observations
are new. All eight Qwen untouched-list views and all eight views in every
Mistral block were also generated after the specification. Thus every block
evaluates the claim against a preassigned disjoint target rather than relabeling
an already examined inclusive result.

\subsection{View treatments and finite estimands}

The ordered bank is
\{canonical, 13, 17, 19, 23, 29, 31, 41\}. The two-view probe
$P_2=\{\text{canonical},13\}$ remains fixed throughout. Probe-size diagnostics
$P_3$ and $P_4$ add 17 and then 19. The fully disjoint four-view target is
$A_4=\{23,29,31,41\}$.

For the controlled overlap treatment, we correlate $V_2(P_2)$ with a four-view
target $V_4(T)$ while changing only the number of reused probe observations:

\begin{itemize}
  \item \textbf{overlap 0:} the single target $T=A_4$;
  \item \textbf{overlap 1:} the mean over the eight targets containing one
  probe view and three views from $A_4$;
  \item \textbf{overlap 2:} the mean over the six targets containing both
  probe views and two views from $A_4$.
\end{itemize}

Every target combination is recomputed and then averaged \emph{within query}.
The eight or six combinations are repeated measurements of the same query, not
additional sampling units. Across-query Spearman correlation is calculated only
after this within-query aggregation. The probe-size analysis similarly
correlates $V(P_k)$, $k\in\{2,3,4\}$, against the unchanged disjoint target
$V_4(A_4)$. A ranking-only analogue,
$1-\text{mean pairwise Kendall }\tau$, checks that the overlap pattern is not
specific to exposure-vector variance.

\subsection{Models, prompts, and deterministic inference}

We use revision-pinned Qwen2.5-7B-Instruct \cite{qwen25report} and
Mistral-7B-Instruct-v0.3 \cite{jiang2023mistral}. The second family was selected
by a predeclared feasibility, licensing, constrained-decoding, and download
tie-break rule; no result entered the choice. Candidate labels A--T are attached
to item identity before serialization, so a permutation changes position but
not identity. Candidate and history text are independently truncated to 64
tokens, and the model must emit all 20 distinct labels.

Inference uses BF16 on one NVIDIA RTX 5090, SDPA, deterministic algorithms,
TF32 disabled, batch size 1, and a 20-step greedy argmax decoder restricted to
unused audited single-token labels A--T. Sampling and output repair are
disabled. Runtime records bind the actual model and input device, precision,
token counts, latency, and peak memory for every block; all blocks ran on
\texttt{cuda:0} with no CPU fallback.

\subsection{Controlled and untouched candidate strata}

The two strata answer different questions and are never pooled
(Table~\ref{tab:strata}). The controlled stratum contains 1,113 MovieLens-1M
queries \cite{harper2015movielens} and 339 KuaiRec queries
\cite{gao2022kuairec}. Each fixed 20-item list includes relevant head and tail
candidates, making a relevance-proportional head/tail exposure target
observable. This mechanism-isolating construction supports signed offline
intervention analysis, but it is not presented as a natural deployment
distribution.

For realism, an outcome-blind hash rule selects 300 sealed users per dataset and
preserves each retriever's untouched top-20 list. These lists provide a more
natural candidate distribution for target-free perturbation-instability
validity. They do not validate the controlled relevance/head-tail target, signed
benefit, or a fairness claim.

\begin{table}[t]
  \centering
  \caption{The two analysis strata have deliberately different evidential
  roles. Signed outcomes are computed only where the offline target is
  observable.}
  \label{tab:strata}
  \small
  \begin{tabular}{@{}p{.18\columnwidth}p{.29\columnwidth}p{.40\columnwidth}@{}}
    \toprule
    Stratum & Candidate list & Supported role \\
    \midrule
    Controlled & Fixed 20-item relevance/head--tail construction &
    Disjoint validity; signed $B_F$; NDCG; routing cost \\
    Untouched & Retriever top 20, hash-selected users &
    Target-free overlap and instability validity only \\
    \bottomrule
  \end{tabular}
\end{table}

Table~\ref{tab:execution} summarizes which rankings already existed and which
were newly generated. ``4+4'' denotes four existing Qwen discovery-side
measurements plus four newly generated target views; earlier inclusive evidence
is not treated as prospective support.

\begin{table}[t]
  \centering
  \caption{Prospective execution blocks. Analysis uses eight views per query;
  ``4+4'' means four existing discovery-side views plus four newly generated
  target views. ML/KR abbreviate MovieLens/KuaiRec.}
  \label{tab:execution}
  \small
  \begin{tabular}{llrrl}
    \toprule
    Model & Stratum & Queries & New rows & Generation \\
    \midrule
    Qwen & ML controlled & 1113 & 4452 & 4+4 \\
    Qwen & KR controlled & 339  & 1356 & 4+4 \\
    Qwen & ML untouched  & 300  & 2400 & all new \\
    Qwen & KR untouched  & 300  & 2400 & all new \\
    Mistral & ML controlled & 1113 & 8904 & all new \\
    Mistral & KR controlled & 339  & 2712 & all new \\
    Mistral & ML untouched  & 300  & 2400 & all new \\
    Mistral & KR untouched  & 300  & 2400 & all new \\
    \bottomrule
  \end{tabular}
\end{table}

\subsection{Signed outcomes and explicit baselines}

On controlled lists, $\loss$ is the maximum absolute head/tail exposure
deviation from a relevance-proportional target. Borda aggregation uses the
disjoint $A_4$ rankings and
\begin{equation}
 \benefit=\loss(\mathrm{canonical})-\loss(\mathrm{Borda}(A_4)),
\end{equation}
so positive $\benefit$ is improvement. We also report the $\ell_2$ exposure
movement, $|\benefit|$, and $\Pr(\benefit>0)$. The signed convention prevents
positive-part summaries from hiding harmful aggregation.

Measurement baselines make each interpretation explicit. The naive inclusive
baseline validates $V_2$ against a target that reuses both probe views; the
primary comparison instead uses the fully disjoint $A_4$. The ranking-only analogue
changes the instability representation, and $P_2/P_3/P_4$ vary probe size while
holding $A_4$ fixed. Canonical entropy and top-two margin are one-call
uncertainty controls rather than permutation diagnostics.

We additionally evaluate fixed-fraction routing as a downstream cost--utility
test. At fixed route fractions
$r\in\{0.10,0.25,0.50\}$, queries are ordered by $V_2$, canonical entropy,
canonical margin, or a deterministic query-hash random score. The $V_2$ policy
spends two mandatory probe calls and acquires four target-view calls when routed, for
$2+4r$ calls on average. Entropy, margin, and random policies observe only the
canonical ranking before routing and therefore cost $1+4r$. This one-call
advantage is structural, not a tuned empirical benefit. Canonical-only and
always-aggregate endpoints bracket the policies. We report actual calls, input
and output tokens, decoder invocations, latency, $\loss$, and NDCG@10. A point
is strictly dominated only if another has no higher measured calls, no worse
$\loss$ and NDCG, and at least one strict improvement.

\subsection{Statistical procedure and reproducibility}

Each query (user) is the independent sampling unit within a model--dataset
block. All uncertainty uses 1,000 paired query bootstraps (pre-specified seeds
2401--3400); each replicate resamples queries and recomputes target
combinations, correlations, and paired contrasts. The confirmatory family
contains eight one-sided tests: overlap-2 minus overlap-0 and $P_4$ minus $P_2$
for each of four controlled blocks. Holm correction is applied across this
family. All other intervals and tests are descriptive, remain block-specific,
and are interpreted through effect size rather than significance alone.
Five-fold held-out prediction and routing analyses keep folds or route fractions
fixed; no threshold is selected against prospective outcomes.

The anonymous supplement supports reproduction from the packaged ranking JSONL
and Parquet records through per-query metrics, tables, and
Figure~\ref{fig:results}. Reviewers can reproduce reported values without
rerunning either language model; inference code, fixed configurations, device
records, and data-acquisition boundaries are included separately.

\section{Results}

\subsection{Observation reuse changes the validity conclusion}

\begin{figure*}[t]
  \centering
  \includegraphics[width=\textwidth]{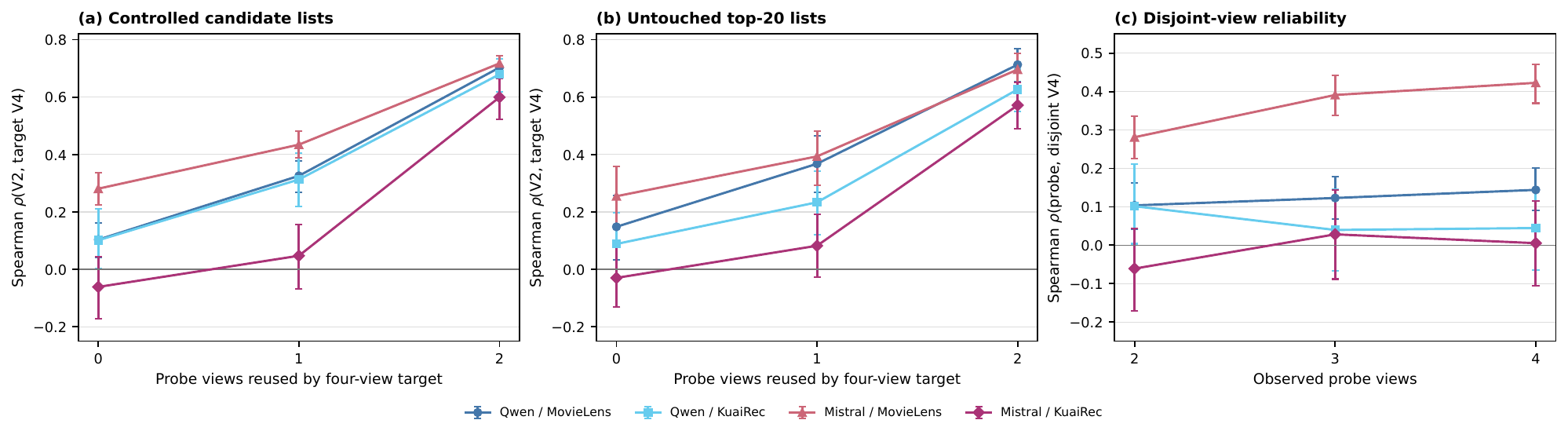}
  \Description{Three panels report Spearman correlations with 95 percent
  bootstrap intervals. On controlled and untouched lists, overlap with probe
  views strongly increases apparent validity. A separate panel shows that
  increasing the probe from two to four views improves disjoint reliability
  clearly only for Mistral on MovieLens.}
  \caption{Prospective validity for the unseen-view estimand. Error bars are
  95\% paired query-bootstrap intervals. Reusing probe observations strongly
  increases association on controlled lists (a) and on the target-free
  untouched-list sensitivity (b). Increasing probe calls improves disjoint-view
  reliability clearly only for Mistral/MovieLens (c).}
  \label{fig:results}
\end{figure*}

Figure~\ref{fig:results}(a) and Table~\ref{tab:primary} show the primary result.
The fully disjoint association $\rho_0$ is modest and heterogeneous
($-0.061$--$0.281$), whereas the two-reuse association $\rho_2$ is
$0.600$--$0.718$. The paired contrast
$\Delta_{2-0}=\rho_2-\rho_0$ is $0.436$--$0.661$ across the four controlled
blocks. Every 95\% interval excludes zero, all four Holm-adjusted $p$ values are
0.008, and the median contrast is 0.589. The claim rests on the size and
replication of this contrast, not on significance alone.

The overlap direction replicates under the untouched candidate distribution.
On the four untouched top-20 blocks, $\Delta_{2-0}$ is 0.565, 0.538, 0.441, and 0.601,
and every descriptive bootstrap interval is positive
(Figure~\ref{fig:results}(b)). This sensitivity supports the target-free
measurement conclusion only; no signed head/tail outcome is defined on these
lists.

The same pattern appears with the ranking-only diagnostic. Its controlled
$\rho_0$ ranges from $-0.113$ to $0.237$, compared with
$0.563$--$0.702$ for $\rho_2$. Removing the explicit probe-pair distance from
the inclusive four-view statistic reduces $\rho_2$ to $0.167$--$0.509$ but does
not eliminate it. Equation~\ref{eq:nested} explains why: cross terms still
contain probe observations. Deleting one self-distance is therefore not
equivalent to evaluation against the fully disjoint four-view target.

The evaluation decision changes in three of four controlled blocks: a reused-
view target exceeds $\rho=.30$, whereas the two-view probe remains below $.20$
when evaluated against the fully disjoint four-view target. The empirical effect
size is specific to this prospectively fixed probe and view bank.

\begin{table*}[t]
  \centering
  \caption{Controlled-list overlap contrast and downstream separation.
  $\rho_0/\rho_2$ use zero/two shared views; $\rho_{move}$ and
  $\rho_{signed}$ correlate $V_2$ with exposure movement and signed benefit.
  Positive $\overline{B_F}$ means the disjoint aggregation improves the offline
  target on average. CIs are 95\% paired query-bootstrap intervals; reported
  overlap $p$ values are Holm-adjusted within the eight-test confirmatory
  family. Other entries are descriptive.}
  \label{tab:primary}
  \resizebox{\textwidth}{!}{\input{tables/primary_results.tex}}
\end{table*}

\subsection{More probe views do not universally restore disjoint reliability}

Figure~\ref{fig:results}(c) shows heterogeneous gains from additional probe
views.
For Mistral/MovieLens, disjoint reliability rises from 0.281 at $P_2$ to 0.423
at $P_4$; the paired gain is 0.142 [0.096, 0.188], Holm $p=.008$. Qwen/MovieLens
rises only 0.040 [-0.011, 0.094]. Qwen/KuaiRec is non-monotone (0.102, 0.040,
0.045), and Mistral/KuaiRec remains near zero. Thus the measurement cost needed
for a reliable diagnostic depends on model and data.

Five-fold held-out prediction gives the same boundary. Adding $V(P_4)$ beyond
$V(P_2)$ raises held-out $R^2$ for the disjoint target by 0.036 [0.013, 0.060]
for Qwen/MovieLens and 0.206 [0.159, 0.248] for Mistral/MovieLens, but by only
$-0.002$ [$-0.023$, 0.017] and $-0.011$ [$-0.021$, $-0.001$] on the two
KuaiRec blocks. The top-quartile query sets selected by the two probes have
Jaccard overlap only 0.13--0.24. Extra views can change both predictive value
and which queries appear unstable; they are not merely a lower-noise version of
the same fixed gate.

\subsection{Predicting movement is not predicting signed benefit}

Across all four controlled blocks, $V_2$ correlates positively with the
canonical-to-$A_4$ aggregation movement ($\rho=0.142$--$0.426$; every interval
above zero). It does not supply stable directional information:
$\rho(V_2,B_F)$ is 0.025, -0.069, 0.034, and -0.131. Three intervals cross zero;
the fourth is negative. Mean benefit is positive on MovieLens for both models
(0.0248/0.0317) and negative on KuaiRec (-0.0450/-0.0329). This matches the
non-identifiability proposition: movement magnitude can be predictable while
direction varies with the external target.

\subsection{Diagnostic validity does not settle the routing decision}

At measured cost, 7 of 12 $V_2$ routing points are strictly dominated by an
endpoint or a no-higher-call entropy, margin, or random point on target loss and
NDCG. The remaining 5 points are nondominated (Table~\ref{tab:dominance}). The
pattern is heterogeneous: all three Qwen/KuaiRec points are dominated, whereas
two points remain nondominated for each MovieLens model block and one remains
for Mistral/KuaiRec. Thus neither uniform superiority nor uniform failure is
supported. The narrower operational conclusion is that reused-view association
cannot justify routing, and even a disjoint-view diagnostic must be evaluated
against signed value and actual cost.

\begin{table}[t]
  \centering
  \caption{Strictly dominated $V_2$ routing points among the three registered
  route fractions. Dominance uses measured calls, target loss, and NDCG; these
  are descriptive fixed-policy comparisons.}
  \label{tab:dominance}
  \small
  \begin{tabular}{lcc}
    \toprule
    Block & Dominated & Nondominated \\
    \midrule
    Qwen / MovieLens    & 1 & 2 \\
    Qwen / KuaiRec      & 3 & 0 \\
    Mistral / MovieLens & 1 & 2 \\
    Mistral / KuaiRec   & 2 & 1 \\
    \midrule
    Total                & 7 & 5 \\
    \bottomrule
  \end{tabular}
\end{table}

\section{Implications for Diagnostic Evaluation}

The results motivate a four-stage evaluation for diagnostics whose intended estimand
concerns unseen perturbation behavior. First, define the perturbation population
and the finite probe. Second, test fresh-view predictive information using
validation-target observations disjoint from those used to construct the probe.
Third, if the diagnostic triggers an action, separately evaluate signed task
utility; association with response magnitude is not association with benefit.
Fourth, compare the induced policy with simple uncertainty gates and fixed
endpoints at measured calls and outcomes rather than nominal threshold labels.
We use the fixed-fraction routing comparison to test this final transition from
diagnostic validity to a cost--utility policy.

This observation-disjoint condition is specific to estimands about unseen draws
from the same perturbation process. A consistency score evaluated against an
external relevance label, ranking-quality judgment, or attack outcome has a
different target and may require a different design.

The overlap curve identifies the effect of observation reuse rather than
modeling a literal deployment procedure. Its three overlap levels hold the
probe, target size, and query fixed while varying only probe--target reuse.
The remaining 0.167--0.509 association after removing the explicit probe-pair
term shows why deleting a single self-distance is insufficient: inclusive
finite-view variance still contains cross terms involving probe observations.
Fully disjoint target observations remove this direct mechanical channel
without assuming that the underlying rankings are independent.

The same estimand discipline may apply to perturbation diagnostics interpreted
as evidence about unseen draws from their own process. The algebra identifies
reuse but not association across new draws, which remains model- and
data-dependent. Our experiments cover deterministic permutations only; they do
not establish the same pattern for stochastic decoding, prompt ensembles,
repeated retrieval, test-time augmentation, or multi-agent voting.

\section{Related Work and Distinction}

\subsection{Classical statistical ancestry}

Part--whole correlation and mathematical coupling have a classical statistical
lineage \cite{henrysson1963itemtotal,archie1981coupling,blance2005coupling}.
They establish that variables calculated from shared measurements can be
associated for algebraic reasons. Circular analysis warns more broadly against
dependent selection and analysis \cite{kriegeskorte2009circular}, but our
prospectively fixed probe is not an exact instance of selective double dipping.
Our paper-specific contribution is to test how this classical dependence
changes validation of a finite perturbation diagnostic with an unseen-view
target, using prospective control of probe--target observation reuse.

ACES is a recent structural held-out analogue: it withholds one generated test
and measures that test's discriminative agreement with code rankings induced by
the remaining tests \cite{sun2026aces}. Its target is generated-test quality in
code selection. The present study instead measures deterministic ranking-view
instability and directly varies zero/one/two observation overlap.

\subsection{Order sensitivity, mitigation, and adaptive computation}

Permutation Self-Consistency aggregates rankings and validates the result
against external ranking quality \cite{tang2024permutation}. RISE evaluates and
mitigates position bias in recommendation \cite{bito2025position}; Self-Sorting
adds selection-time reranking \cite{do2026selfsorting}; CapCal calibrates
content-agnostic positional priors \cite{lv2026capcal}; and DebiasFirst,
InvariRank, and Set-LLM pursue training-time or architectural robustness
\cite{qiao2026debiasfirst,bito2026invarirank,egressy2025setllm}. Their targets
are effectiveness, reduced positional bias, or invariant behavior, rather than
validation of a small diagnostic against unseen views.

RoToR routes mixed order-sensitive and order-invariant listwise inputs
\cite{yoon2025rotor}. AcuRank adapts listwise computation from an evolving
TrueSkill relevance posterior and evaluates an accuracy--efficiency trade-off
\cite{yoon2025acurank}. These are the closest selective-routing and
adaptive-computation precedents, respectively. Our fixed-fraction comparison
instead tests whether a pre-specified permutation diagnostic yields a useful
cost--utility policy; it does not emulate their distinct states or stopping
rules.

\subsection{Consistency as an evaluation object or predictive signal}

Pres et al. frame self-consistency as relationships among a model's responses
across inputs and as an optimization objective \cite{pres2026selfconsistency}.
Our experiment instead uses a finite probe, a disjoint target, and controlled
observation overlap. Bito et al. target preference consistency,
ranking effectiveness, and marginal exposure \cite{bito2026preference}; Ni et
al. use self-consistency to predict externally labeled ranking quality
\cite{ni2026qpp}; and Zhang et al. predict attack vulnerability
\cite{zhang2026attack}. Multiple-choice evaluation likewise compares
consistency with externally labeled accuracy \cite{li2024mcqvalidity}.

These targets are substantively different from unseen perturbation behavior and
can legitimately support different validation designs. Our distinction is the
combination of a finite deterministic-view probe, a fresh-view estimand,
prospective control of observation reuse, within-query treatment of view
combinations, and separate signed-value and measured-cost tests. This
distinguishes our fresh-view estimand and controlled overlap manipulation from
external-label and optimization objectives.

\section{Limitations and Failure Boundary}

The algebra applies to the implemented population finite-view variance, while
the empirical effect sizes are specific to one fixed eight-view generator and
the canonical/seed-13 primary probe. We do not sample random probe banks from an
abstract permutation population. With only eight views, the largest probe that
still leaves a four-view disjoint target has size four, and additional views do
not restore reliability in every observed block.

The empirical study covers two pinned 7B instruction-model families, two
recommendation datasets, fixed 20-item lists, and deterministic constrained
decoding. It does not test larger models, passage retrieval, stochastic
decoding, alternative view generators, or different list lengths. The four
model--dataset blocks are reported separately rather than treated as a random
sample of models or domains.

Controlled candidates make a signed relevance-proportional head/tail exposure
target observable; they are not ordinary deployment lists. Untouched-list
results support only target-free instability validity and do not repair the
external-validity limit of the signed analysis. The target is an experimental
popularity head/tail construct, not protected-group, provider, or individual
fairness, and no group label enters the diagnostic. All outcomes are offline;
we make no online, causal, or welfare claim. Finally, five routing points remain
nondominated, so the evidence supports a setting-specific validity boundary
rather than a universal judgment about permutation instability.

\section{Conclusion}

When a finite perturbation diagnostic is interpreted as evidence about unseen
perturbation behavior, a validation target that reuses probe observations does
not isolate fresh-view predictive information. In our prospective study,
fully disjoint evaluation materially changes that conclusion across two model
families and two datasets. More probe views do not repair reliability uniformly;
fresh-view predictiveness, signed intervention value, and measured-cost policy
utility remain separate validity questions. Perturbation diagnostics should
therefore be validated at the level of the claim they are intended to support.

\section{Ethics Statement}

This is an offline study of public recommendation datasets and pinned open
model checkpoints. We make no claim about protected attributes, individual
fairness, provider fairness, or online welfare. The head/tail partition is an
experimental popularity construct, and target labels are used only after
inference. Compute, failed and heterogeneous results, fixed run identities, and
actual-device execution are documented in the anonymous supplement.

Misinterpreting shared-view association as fresh-view reliability could waste
inference, motivate poorly justified routing or fallback decisions, or create
unwarranted confidence in behavior under unseen perturbations. For diagnostics
intended to support unseen-view claims, mitigations are observation-disjoint
validation, a separate signed task-utility test before intervention, and
actual-cost rather than nominal-policy accounting. Any deployment use should
additionally monitor the diagnostic and triggered actions in its specific
operating environment.

\section{Generative AI Disclosure}

GPTImage2 rendered Figure~\ref{fig:validity-chain} from an author-specified
technical prompt. The authors identified and corrected one duplicated probe
view and verified every final label, seed, branch, and offline-only data path
against the pre-specified design. The tool did not choose the hypothesis,
analyze results, or determine claims. Figure~\ref{fig:results} was generated
deterministically from the packaged result CSV files.

\bibliographystyle{ACM-Reference-Format}
\bibliography{references}

\end{document}

%% file: tables/primary_results.tex
\begin{tabular}{lrrrrrrr}
\toprule
Block & $\rho_0$ & $\rho_2$ & $\Delta_{2-0}$ [95\% CI] & Holm $p$ & $\rho_{move}$ & $\rho_{signed}$ & $\overline{B_F}$ \\
\midrule
Qwen / ML & 0.104 & 0.703 & 0.600 [0.548, 0.653] & 0.008 & 0.426 & 0.025 & 0.025 \\
Qwen / KR & 0.102 & 0.680 & 0.578 [0.485, 0.674] & 0.008 & 0.343 & -0.069 & -0.045 \\
Mistral / ML & 0.281 & 0.718 & 0.436 [0.391, 0.480] & 0.008 & 0.380 & 0.034 & 0.032 \\
Mistral / KR & -0.061 & 0.600 & 0.661 [0.559, 0.766] & 0.008 & 0.142 & -0.131 & -0.033 \\
\bottomrule
\end{tabular}